%% file: arxiv_clean.tex
\documentclass[letterpaper]{article}
\usepackage{natbib,alifeconf}
\usepackage{amsmath}
\usepackage{graphicx,subfigure}
\usepackage{tikz}
\usetikzlibrary{arrows}
\usetikzlibrary{positioning}
\usepackage{appendix}

\title{Towards designing artificial universes for artificial agents under \\interaction closure}
\author{Martin Biehl, Christoph Salge \and Daniel Polani\\
\mbox{}\\
University of Hertfordshire, Hertfordshire, UK\\
m.biehl@herts.ac.uk}
\begin{document}

\maketitle

\begin{abstract}
 We are interested in designing artificial universes for artificial agents. We view artificial agents as networks of high-level processes on top of of a low-level detailed-description system. We require that the high-level processes have some intrinsic explanatory power and we introduce an extension of \emph{informational closure} namely \emph{interaction closure} to capture this. Then we derive a method to design artificial universes in the form of finite Markov chains which exhibit high-level processes that satisfy the property of interaction closure. We also investigate control or information transfer which we see as an building block for networks representing artificial agents.
 
\end{abstract}

\section{Introduction}
\input{arxiv_intro_aphys.tex}

\section{Related work}

In general, artificial agents have been studied using information theoretical concepts by several authors (e.g. \citet{klyubin_organization_2004,lungarella_methods_2005,bertschinger_autonomy_2008,williams_information_2010,zahedi_quantifying_2013}). Of those authors many also employ Bayesian networks and specifically the perception-action loop \citep{klyubin_organization_2004,bertschinger_autonomy_2008,zahedi_higher_2009}). The perception-action loop is a Bayesian network describing the causal relations between four stochastic processes representing environment, sensor, actuator, and memory (of the agent) states respectively. In these papers the perception-action loop is not seen as a network of high-level processes in our sense since the interactions between the four processes are direct and not mediated by an underlying process. 

As already mentioned our notion of interaction closure is an extension of the concept of informational closure introduced by \citet{bertschinger_information_2006}. The main difference is that we define interaction closure between two processes with respect to a third (the underlying one) while the original notion concerns closure of one process with respect to another only. We also use a stronger version of informational closure. 

Conditions on underlying processes to exhibit ``independence'' of a high-level process from an underlying one have been studied for Markov chains at least since \citet{kemeny_finite_1976}. They study \emph{lumpability} which requires that the high-level process is itself a Markov process. Research in this direction has been extended in \citet{gornerup_method_2008,jacobi_spectral_2009}. Very recently lumpability has been shown to be implied by informational closure by \citet{pfante_comparison_2013}. In this work various other level structure measures have also been thoroughly investigated. Interactional versions were not studied though.

Our notion of apparent control or information transfer is studied in the context of distributed computation in great detail by \citet{lizier_framework_2014}. It is argued there that information transfer (measured in the same way as here) is one of three ingredients needed for computation the other two being information storage and information modification. Investigations into the computational capabilities of dynamical systems have a long history (e.g. \citet{langton_computation_1990,mitchell_revisiting_1993} and see \citet{lizier_framework_2014} for more). As far as we know, the focus there has not been on the implications of computation occurring on a high-level for the underlying process.

\section{Formal concepts}

\subsection{Artificial universe}
We start by representing an isolated system (referred to as an \textit{artificial universe} or the \textit{underlying process} in the following) by a finite Markov chain\footnote{We choose the index set $I$ as the integers and initialize the process in its stationary distribution at $t=0$.} $\{X_t\}_{t \in I}$ on state space $\mathcal{X}$ defined by the time-homogenous transition kernel (or Markov matrix) $P:=p(X'|X):=(p_{x'x})$ with 
\begin{equation}
 p_{x'x} := p(x'|x) := Pr(X_{t+1}=x'|X_t=x).
\end{equation} 
Our assumption is that the isolated system should be Markov, as there is no external storage of information about past states. Choosing finiteness and time discreteness is done to reduce technical issues and improve clarity of the concepts, for the same reason we restrict ourselves to the stationary case in this treatment. Stationarity may often be a valid approximation for some time interval.

\subsection{High level processes}
We call a random process $\{Y_t\}_{t \in I}$ on state space $\mathcal{Y}$ a \textit{high-level process} of $\{X_t\}_{t \in I}$, if $Y_t$ is dependent only on $X_t$ via a transition matrix $\Pi^Y=(\pi^Y_{yx})$ defined by 
\begin{equation}
\label{derprocess}
 \pi^Y_{yx} := \pi^Y(y|x) := Pr(Y_t=y|X_t=x).
\end{equation} 
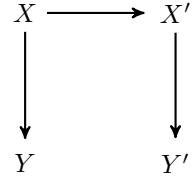
\begin{figure}
\begin{center}
  \begin{tikzpicture}
    [->,>=stealth',shorten >=2pt,auto,node distance=2cm,
    thick,main node/.style={font=\sffamily\normalsize\bfseries}]

    \node[main node] (1) [] {$X$};
    \node[main node] (2) [right of=1] {$X'$};
    \node[main node] (3) [below of=1] {$Y$};
    \node[main node] (4) [below of=2] {$Y'$};

    \path[every node/.style={font=\sffamily\small}]
      (1) edge node {} (2)
      (1) edge node {} (3)
      (2) edge node {} (4)

      ;
  \end{tikzpicture}
  \caption{Bayesian network representing one time step of the relationship of the underlying process $\{X_t\}_{t \in I}$ and a high-level process $\{Y_t\}_{t \in I}$. The primed random variables represent the process state one time step after the not primed ones.}
  \label{fig:derivedprocess}
\end{center}
\end{figure}
Note that the transitions $\pi_{xy}^Y$ are independent of time. See Fig. \ref{fig:derivedprocess} for the corresponding causal Bayesian network \footnote{Following \citet{pearl_causality:_2000} we only draw arrows for causal interactions. Our measures on the other hand are all purely observational.}. 
We also define the Bayesian inverse:
\begin{equation}
\label{binverse}
 \pi^{Y\dagger}_{xy} = \begin{cases} 
0 &  \text{if} \;\pi^Y(y|x)=0 \\ 
\frac{\pi^Y(y|x)\;p(x)}{p(y)} &  \text{else}
\end{cases}
\end{equation} 
where $p(x)$ is the stationary distribution. For a detailed investigation of high-level processes see the work of \citet{pfante_comparison_2013}. 

We also explicitly mention the deterministic case. Call a random process $\{Y_t\}_{t \in I}$ on state space $\mathcal{Y}$ a \textit{deterministic high-level process of} $\{X_t\}_{t \in I}$, if $Y_t = f^Y(X_t)$ for some function $f:\mathcal{X}\rightarrow {\mathcal{Y}}$ we can represent such a function $f^Y$ by a matrix $\Pi^Y=(\pi^Y_{yx})$ defined by 
\begin{equation}
\label{detderprocess}
 \pi^Y_{yx} := \pi^Y(y|x)= \delta_{f^Y(x)}(y):=\begin{cases} 
1 &  \text{if} \;f^Y(x)=y \\ 
0 &  \text{else}
\end{cases}
\end{equation} 
Again transitions are independent of time.
The Bayesian inverse reduces to:
\begin{equation}
 \pi^{Y\dagger}_{xy} = \begin{cases} 
0 &  \text{if} \;x \notin (f^Y)^{-1}(y) \\ 
\frac{p(x)}{p(y)} &  \text{else}
\end{cases}
\end{equation}
where 
\begin{equation}
 p(y) = \sum_{x \in (f^Y)^{-1}(y)} p(x).
\end{equation}

\subsection{Weak and strong informational closure}
Informational closure was introduced by \citet{bertschinger_information_2006} to formalize the idea of closure known from systems theory (see references ibid.) within the framework of information theory. Loosely speaking, closure is attained by a system if it can be described without reference to the environment that it is part of \citep{bertschinger_information_2006}. We will distinguish between a weak and a strong form of informational closure. For a high-level process $\{Y_t\}_{t \in I}$ and underlying process $\{X_t\}_{t \in I}$ (Fig. \ref{fig:derivedprocess}) \textit{weak informational closure} is defined by (see \citet{pfante_comparison_2013}):
\begin{equation}
 I(Y':X|Y)=0
 \label{closure}
\end{equation} 
where $I(Y':X|Y)$ is the conditional mutual information. The conditional mutual information for three arbitrary random variables $X,Y,Z$ is defined by
\begin{equation}
 I(X:Y|Z) = \sum_{z} p(z) \sum_{x,y} p(x,y|z) \log \frac{p(x,y|z)}{p(x|z) p(y|z)}.
\end{equation} 
Intuitively one can read this as the amount of extra information $Y$ contains about $X$ that is not already in $Z$. So informational closure (Eq. \ref{closure}) requires that the current high-level process state $Y$ is as predictive with respect to the next high-level process state $Y'$ as the current underlying process state $X$. Note that this condition can be made stronger by requiring that $Y$ is even as predictive of $Y'$ as the \textit{next} underlying process state $X'$. This is expressed by what we will call \textit{strong informational closure}:
\begin{equation}
 I(Y':X'|Y)=0.
\end{equation} 
It follows from the definition of high-level processes that strong informational closure implies weak informational closure (see Appendix A). Note that none of these conditions actually change the causal structure of the Bayesian network. 

\subsection{Interaction closure}
We now extend the concept of strong informational closure to two high-level processes. Given two high-level processes $\{Y_t\}_{t \in I}$ and $\{Z_t\}_{t \in I}$ and an underlying process $\{X_t\}_{t \in I}$, we say that we have \textit{strong interaction closure} \textit{from} $\{Y_t\}_{t \in I}$ \textit{to} $\{Z_t\}_{t \in I}$ if 
\begin{equation}
 I(Z':X'|Y) = 0.
 \label{mtrongaclosure}
\end{equation}
This implies (see Appendix A) the \textit{weak interaction closure}:
\begin{equation}
 I(Z':X|Y) = 0,
 \label{weakaclosure}
\end{equation}
and 
\begin{equation}
 I(Z':Y) = I(Z':X)=I(Z':X').
\end{equation}
The idea behind interaction closure is, that the states of one process are as predictive of the other's next states as the states (current or next respectively) of the underlying process.  

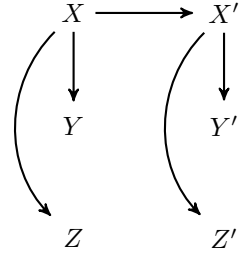
\begin{figure}
\begin{center}
  \begin{tikzpicture}
    [->,>=stealth',shorten >=2pt,auto,node distance=2cm,
    thick,main node/.style={font=\sffamily\normalsize\bfseries}]

    \node[main node] (1) [] {$X$};
    \node[main node] (2) [right of=1] {$X'$};
    \node[main node] (3) [below of=1, node distance=1.5cm] {$Y$};
    \node[main node] (4) [below of=2, node distance=1.5cm] {$Y'$};
    \node[main node] (5) [below of=3, node distance=1.5cm] {$Z$};
    \node[main node] (6) [below of=4, node distance=1.5cm] {$Z'$};

    \path[every node/.style={font=\sffamily\small}]
      (1) edge node {} (2)
      (1) edge node {} (3)
      (2) edge node {} (4)
      (1) edge[bend right=45] node {} (5)
      (2) edge[bend right=45] node {} (6)
      ;
  \end{tikzpicture}
  \caption{Bayesian network representing one time step of an underlying process $\{X_t\}_{t \in I}$ and high-level processes $\{Y_t\}_{t \in I}$ and $\{Z_t\}_{t \in I}$.}
  \label{fig:actorsensor}
\end{center}
\end{figure}

\subsection{Apparent control}
In order to measure in how far one high-level process $\{Y_t\}_{t \in I}$ appears\footnote{Actual control would require a direct causal influence.} to control another high-level process $\{Z_t\}_{t \in I}$ we use the one-step transfer entropy 
\begin{equation}
 I(Z':Y|Z)
\end{equation} 
\citep{schreiber_measuring_2000}. Transfer entropy has been shown to be a measure of controllability by \citet{touchette_information-theoretic_2004}. Here we say that \textit{$\{Y_t\}_{t \in I}$ appears to control $\{Z_t\}_{t \in I}$} if 
\begin{equation}
 I(Z':Y|Z) >0.
\end{equation}
We call this apparent control because in our case the random variable $Y$ is part of a high-level process, and does not represent a true controller. The cause of the dynamics of $\{Z_t\}_{t \in I}$ remains $\{X_t\}_{t \in I}$.

We could also use the term ``information transfer'' as in \citet{lizier_framework_2014} to put more emphasis on the relation to computation, but as control was the first thing we had in mind we stick to it in this publication\footnote{Also, we don't want to discuss here what ``apparent computation'' would be.}.

Note that strong interaction closure does not imply apparent control, e.g. let $\{Y_t\}_{t \in I} = \{Z_t\}_{t \in I}$ then according to the definitions strong interaction closure implies that apparent control is zero. This is due to the fact that apparent control is based on non-causal transfer entropy and therefore a process can never (apparently) control itself. 

We also use the definition of perfect apparent control \citep{touchette_information-theoretic_2004} to express the case where apparent control is maximal. 

\textit{Perfect apparent control} means for all initial states $z \in \mathcal{Z}$ and all final states $z' \in \mathcal{Z}$ there exists a state $y \in \mathcal{Y}$ such that 
\begin{align}
p(z' |z, y) &= 1.
\end{align}
Then $I(Z':Y|Z) = H(Z'|Z)$ i.e. the transfer entropy attains its maximum value. 

\section{Results}
\input{highlevel.tex}
\subsection{Implications of interaction closure}
We now present the implications of strong interaction closure for the underlying process. In order to keep the necessary technical terminology to a minimum we make a few more assumptions which lead to stronger results. 

In the following we will denote the process from which the interaction closure ``originates''  by $\{S_t\}_{t \in I}$ and the ``receiving'' one by $\{M_t\}_{t \in I}$. 
This is done to conform to an interpretation as a sensor that (apparently) writes or transfers information to a memory. In this case strong interaction closure reads:
\begin{equation}
 I(M':X'|S) = 0.
\end{equation}
In Appendix B. we show that under strong interaction closure and the two extra assumptions $|\mathcal{M}|=|\mathcal{S}|$ and $\{M_t\}_{t \in I}$ deterministic i.e. 
\begin{equation}
 \pi^M_{mx} = \delta_{f^M(x)}(m)
\end{equation} 
the following hold (see also Fig.\ref{fig:sandm}):

The process $\{S_t\}_{t \in I}$ is also deterministic with respect to $\{X_t\}_{t \in I}$ and we have an associated function $f^S:\mathcal{X}\rightarrow \mathcal{S}$.

Moreover, for each $m' \in \mathcal{M}$
\begin{equation}
 p(m'|x) = \delta_{f^{M'}(x)}(m')
\end{equation} 
for some function $f^{M'}:\mathcal{X}\rightarrow \mathcal{M}$. Also 
for each $m' \in \mathcal{M}$
\begin{equation}
 p(m'|s) = \delta_{g(s)}(m')
\end{equation} 
for some bijective function $g:\mathcal{S}\rightarrow \mathcal{M}$ with $g:=f^M \circ (f^S)^{-1}$.

Furthermore,
\begin{equation}
\label{xGxsupport}
 p(x'|x) = \begin{cases} 
0 &  \text{if} \;x' \notin (f^M)^{-1} \circ f^{M'}(x)\\ 
\geq 0 & \text{else},
\end{cases}
\end{equation} 
and 
\begin{equation}
\label{xGasupport}
 \pi^{S\dagger}(x|s) = \begin{cases} 
0 &  \text{if} \;x \notin (f^{M'})^{-1}\circ g(s)\\ 
\geq 0 & \text{else}.
\end{cases}
\end{equation}

We have thus arrived at a condition on the transition matrix of the artificial universe process from the requirement of strong interaction closure. There are two main things to take away from this. 

\paragraph{The first} is how to construct a transition matrix that obeys strong interaction closure. For this choose a finite set $\mathcal{X}$ with $|\mathcal{X}|=n$. Then take two sets $\mathcal{M}$ and $\mathcal{S}$ with $|\mathcal{M}|=|\mathcal{S}|$ and functions $f^M:\mathcal{X}\rightarrow\mathcal{M}$ and $f^S:\mathcal{X}\rightarrow\mathcal{S}$. Then construct a matrix, split it vertically according to the preimages $(f^S)^{-1}$ and horizontally according to those of $(f^M)^{-1}$ (if for example the first and the last row are part of $(f^M)^{-1}(m)$ make sure to remember they belong to the same block). Make sure that each column sums to one, and note that the entries in each column can only be larger than zero in one block of the preimage of $(f^M)^{-1}$. Here is an example with $\mathcal{X}=\{1,2,3,4,5,6\}$, $\mathcal{M}=\mathcal{S}=\{1,2\}$, $f^M(x) =1$ for $x \leq 3$ else $f^M(x) =2$ and $f^S(1)=f^S(4) =1$ else $f^S(x) =2$:
\begin{equation}
\label{example}
P=\left(
\begin{array}{cccccc}
 \frac{1}{3} & 0 & 0 & \frac{1}{3} & 0 & 0 \\
 \frac{1}{3} & 0 & 0 & \frac{1}{6} & 0 & 0 \\
 \frac{1}{3} & 0 & 0 & \frac{3}{6} & 0 & 0 \\
 0 & \frac{1}{3} & \frac{1}{2} & 0 & \frac{1}{4} & \frac{1}{2} \\
 0 & \frac{1}{3} & \frac{1}{4} & 0 & \frac{1}{2} & 0 \\
 0 & \frac{1}{3} & \frac{1}{4} & 0 & \frac{1}{4} & \frac{1}{2} \\
\end{array}
\right)\end{equation}

\paragraph{The second} is that we have two partitions on the state space $\mathcal{X}$ induced by the two functions $f^M$ and $f^{M'}$. The former, $(f^M)^{-1}$ partitions $\mathcal{X}$ into blocks of states mapped to the same $m \in \mathcal{M}$ at the current time step and we call it the \textit{current partition}. The latter partitions $\mathcal{X}$ into blocks that are mapped to the same $m \in \mathcal{M}$ at the next time step and we call it the \textit{future partition}. Note that as $g$ is bijective we can also view the future partition as induced by $f^S=g^{-1} \circ f^{M'}$ which shows that $s \in \mathcal{S}$ indicates the blocks of the future partition at the current time step. Note that time evolution starting in $(s,m)$ would be $(s,m),(s',g(s)),(s'',g(s')),...$.  Here $s',s'',...$ are determined by the underlying dynamics. 

The relation between the two partitions can take two extreme cases. The first is, when they \textit{coincide} i.e. if for every $m \in \mathcal{M}$ exists $s \in \mathcal{S}$ such that $(f^M)^{-1}(m) \subseteq  (f^{S})^{-1}(s)$ and vice versa. The other extreme case is when they are \textit{orthogonal} i.e. when for every pair $m,s \in \mathcal{M}\times \mathcal{S}$ we have $(f^M)^{-1}(m) \cap (f^{S})^{-1}(s) \neq \emptyset$. 

For coinciding partitions the blocks coincide and each block has unique associated high-level states $s\in \mathcal{S}$ and $m \in\mathcal{M}$. This means given $s$ for a block, $m$ is determined and vice versa. There is then a bijective function $h:\mathcal{S}\rightarrow\mathcal{M}$ which maps the current $s$ to the \textit{current} $m$ ($g$ maps it to $m'$ the next high-level state). We can then write $M=h(S)$ and $S=h^{-1}(M)$, the two processes up to changes of the alphabet identical.

For orthogonal partitions, in every block of the current partition there is at least one element of every block in the future partition. This means by only knowing the block of the current partition i.e. $m \in \mathcal{X}$ does not tell us anything about the current $s$ or the next $m'=g(s)$. 

\subsection{Implications of apparent control and strong interaction closure}
Here we only look at implications for apparent control under the same assumptions as in the last section.

Recall that apparent control is measured in this context by $I(M',S|M)$. 
We then have the current and the future partition of $\mathcal{X}$. We consider the two extreme cases of coinciding partitions and orthogonal partitions. For coinciding partitions, apparent control vanishes. 
To see this recall that we have a the bijective function $h$ (see last section) such that 
\begin{equation}
 I(M',S|M)=I(M',h^{-1}(M)|M)=0.
\end{equation} 
To see this note that the random variable $h^{-1}(M)$ can never contain more information than $M$ itself.

If we look at the orthogonal case we have that for every block of the current partition indicated by $m \in \mathcal{M}$ and every $m' \in \mathcal{M}$ there is an $x \in \mathcal{X}$ with $f^M(x)=m$ and  $f^S(x)=m$ and $g(s)=m'$. But this just implies perfect apparent control, as in this case
\begin{equation}
 p(m'|m,s)=1.
\end{equation} 

So our measure of apparent control varies from $0$ to its maximum $H(M'|M)$ due to the possible relations between the current and future partitions. 

We can also ask whether perfect apparent control implies orthogonal partitions. As we need for every $m,m' \in \mathcal{M}$ an $s \in \mathcal{S}$ with 
\begin{equation}
 p(m'|m,s)=1.
\end{equation}
we can see that in every block of the current partition corresponding to $m$ there must be elements $x$ in the future partition (i.e. $f^S(x)=s$) that lead to each $m'$. Due to strong interaction closure, and $|\mathcal{S}|=|\mathcal{M}|$ we have a one-to-one relation between $m'$ and $s$ given by $g$, so there must be elements $x$ corresponding to each $s$ in each block of the current partition. This means the two partitions are orthogonal. 

In order to construct a transition matrix of a system with a pair of high-level processes, strong interaction closure and perfect apparent control, follow the procedure for constructing the transition matrix for strong interaction closure only. Make sure though that for each $s$ and $m$ there is a state $x \in (f^S)^{-1}(s) \cap (f^M)^{-1}(m)$. For example in the example of the last section with $\cap(s,m):=(f^S)^{-1}(s) \cap (f^M)^{-1}(m)$ we find $\cap(1,1)=\{1\}, \cap(1,2)=\{4\}, \cap(2,1)=\{2,3\}, \cap(2,2)=\{5,6\}$ and thus we have perfect apparent control there. We find also that, as expected, $I(M',S|M)=H(M'|M)= 0.95669$.

\section{Discussion}
We were looking for design principles for artificial universes especially with regard to the capability to contain artificial agents on a higher or macroscopic level. Conceptualizing artificial agents as networks of high-level processes, we focussed on the interaction of two such processes. To formalize the condition that there should be some explanatory power on the macroscopic level we introduced interaction closure as an extension to informational closure. 

We found that if we require interaction closure, equal cardinalities of the high-level processes' state spaces and determinism of the receiving process, the dynamics of the underlying process must respect (see Eqs. \ref{xGxsupport}, \ref{xGasupport}) two partitions of state space \footnote{The partitions also exist and are respected if the receiving process is not deterministic but the cardinality of its set of extreme points is equal to the cardinality of the other process (see Eqs. \ref{xgxgeneral} and \ref{xgsweak2}).}. How the two partitions are related is not determined by interaction closure. In other words, interaction closure does not specify the kind of interaction and requires only that it is closed with respect to the underlying process. To design an underlying process we can then choose the partitions (which induce the two processes) freely and create the transition matrix accordingly (see Results). Considering that we can choose the underlying state space arbitrarily large we expect that a large 
variety of high-level dynamics can be 
implemented in this way.

We also investigated a special kind of interaction, apparent control, between the high-level processes. It can be interpreted as one high-level process controlling the other or as one process transferring information to the other. We identified to extreme cases which occur. The first occurs if the two partitions associated with the interaction closure coincide, the two high-level process are essentially the same, and apparent control vanishes. The second occurs when the two partitions are orthogonal, the two high-level processes are complementary, and control is maximal. Intermediate relations between the partitions would led to intermediate levels of control. 

In the future we want to investigate complete networks of high-level processes that are informationally and interactionally closed. Further interesting measures are the other ingredients of computation, information storage and modification as well as their localized versions \citep{lizier_framework_2014}. These are interesting to us because computation seems relevant for artificial agents. We also want to focus on network structures relevant for artificial agents with metabolisms.

\bibliographystyle{apalike}
\bibliography{../bibliography}

\appendix
\section*{APPENDIX}
\renewcommand{\thesubsection}{\Alph{subsection}}

\subsection{A.}
\label{appstrong}
To see that strong interaction closure implies weak interaction closure (snd therefore strong informational closure implies weak informational closure), note
\begin{align}
\label{doublecmi}
 I(Z':X,X'|Y) &= I(Z':X'|Y) + I(Z':X|X',Y)\\
              &= 0
\end{align}
where the first term on the right vanishes because it represents strong interaction closure and the second term vanishes because $\{X',Y\}$ d-separates $Z'$ and $X$ according to the Bayesian network in Fig. \ref{fig:actorsensor}. In general:
\begin{equation}
 I(Z':X|Y) \leq I(Z':X,X'|Y)
\end{equation} 
and as conditional mutual informations are non-negative, $I(Z':X|Y)=0$ as well, which means we have weak interaction closure. By replacing $Z'$ by $Y'$ the same argument also proves the informational closure. 
For d-separation in the context of Bayesian networks and conditional mutual information see \citet{ay_information_2008}.

To see that strong interaction closure implies 
\begin{equation}
\label{miequals}
 I(Z':Y) = I(Z':X)=I(Z':X')
\end{equation}  
consider
\begin{align}
 I(Z':Y,X) &= I(Z':Y) + I(Z':X|Y)\\
           &= I(Z':X) + I(Z':Y|X).
\end{align} 
In both lines the second terms on the right hand side vanish. In the upper case because this is the requirement of weak interaction closure (which is implied by the strong version) and in the lower equation because $X$ d-separates $Z'$ and $Y$. This gives us the first equality in Eq.\ref{miequals}, the second follows by replacing $X$ by $X'$ and using the same reasoning.

\subsection{B.}
\input{arxiv_proofs2.tex}

\end{document}

%% file: arxiv_intro_aphys.tex
We are interested in designing artificial physics for artificial agents. This paper presents an exploratory step in this direction and also expounds the conceptual and the formal point of view we are taking. In this introduction we give a short overview of our approach and then proceed to formally define the different elements. 

Conceptually, we draw inspiration for our artificial physics and agents from ``real'' physics and living organisms. The artificial agents we have in mind are are minimally represented by networks of ``high-level'' or ``macroscopic'' processes. These high-level processes are derived from the underlying artificial physics. This situation is analogous to viewing living organisms as networks of processes \citep{maturana_autopoiesis_1980} on a meso- or macroscopic scale e.g. proteins or cells, and assuming an underlying physics e.g. elementary particle physics. Formally, we model our artificial physics simply as a univariate finite discrete time Markov process. We choose a univariate process because we do not want to presuppose any structure of the state space of the artificial physics. We also assume there is no downward causation \citep{campbell_downward_1974}. This means that at all times, the high-level processes are causally dependent on the underlying physics. Loosely speaking, this means that the edges (interactions) of the high-level network of processes representing the agent are actually mediated by the low-level process. As we will see, this can formally be modelled using Bayesian networks. 

The final ingredient of our general approach tries to account for the success of doing science on scales larger than elementary particles e.g. atomic physics, chemistry and biology. To take this into account, we require that the high-level processes are as predictive of other high-level processes as the underlying physics itself. In other words, the high-level processes at least appear to be directly causally related. Formally, we achive this by slightly extending the notion of informational closure introduced by \citet{bertschinger_information_2006} to two notions that we will call \textit{weak and strong interaction closure}. Requiring informational closure already puts some constraints on the underlying process \citep{pfante_comparison_2013} and so do interaction closures. 

Within this general setting we here inspect the situation where one high-level process seems to control another one. The idea is that any high-level network that represents an agent needs such a mechanism. Consider for example a sensor that writes its measurement to another process e.g. a memory for further processing. Another interpretation would be that the controlled process is part of the embodiment of the agent and therefore within the sphere of influence of the agent and shielded from the environment. The latter interpretation is related to the notion of embodiment put forward by \citet{porr_inside_2005}. Yet another, more conservative, interpretation would be that the first process simply transfers information to the second. Information transfer is widely seen as an important part of decentralized computation \citep{lizier_framework_2014}. Which in turn may be just what a network of processes representing an agent needs. Formally, we use an information theoretic notion, the transfer entropy \citep{schreiber_measuring_2000}, to quantify (here only apparent) control. Control and transfer entropy have been linked in another context by \citet{touchette_information-theoretic_2004}. 

Note that the mechanism we treat is a requirement we introduce here in addition to interaction closure property. In order to arrive at a complete agent further mechanisms within larger networks are required. This will be investigated in future work.     

The results in this paper show that the requirements of strong interaction closure and control from a pair of high-level processes put strong constraints on the dynamics of the underlying process. To arrive at these constraints we assume the ideal cases of both interaction closure and control. It should be seen as an advantage of the information theoretic measures we employ that they are both ``soft''. This means they can readily be used to quantify also the degrees to which closure and control are present in a system.

%% file: highlevel.tex
\begin{figure}
\begin{center}
  \begin{tikzpicture}
    [->,>=stealth',shorten >=2pt,auto,node distance=2cm,
    thick,main node/.style={font=\sffamily\normalsize\bfseries}]

    \node[main node] (1) [] {$X$};
    \node[main node] (2) [right of=1] {$X'$};
    \node[main node] (3) [below of=1, node distance=1.5cm] {$S$};
    \node[main node] (4) [below of=2, node distance=1.5cm] {$S'$};
    \node[main node] (5) [below of=3, node distance=1.5cm] {$M$};
    \node[main node] (6) [below of=4, node distance=1.5cm] {$M'$};

    \path[every node/.style={font=\sffamily\small}]
      (1) edge node {} (2)
      (1) edge node {} (3)
      (2) edge node {$\delta_{f^{S}}$} (4)
      (1) edge[bend right=45] node [left] {$\delta_{f^{M}}$} (5)
      (2) edge[bend right=45] node [pos=0.8] {} (6)
      (1) edge[dashed, bend right=12.43] node [pos=0.15,right] {$\delta_{f^{M'}}$} (6)
      (3) edge[dashed, bend right=12.43] node[pos=0.5,below] {$\delta_{g}$} (6)
      ;
  \end{tikzpicture}
  \caption{Bayesian network representing one time step of an underlying process $\{X_t\}_{t \in I}$ and high-level processes $\{S_t\}_{t \in I}$ and $\{M_t\}_{t \in I}$. We indicate for the case mentioned in the Result section the mechanisms associated with transitions. Dashed arrows are not part of the Bayesian network (not causal). Note $\delta_{f^S}$ is also associated with $X\rightarrow S$ and $\delta_{f^M}$ also with $X'\rightarrow M'$. This is not indicated due to space limitations.}
  \label{fig:sandm}
\end{center}
\end{figure}
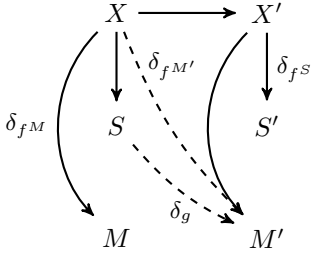

%% file: arxiv_proofs2.tex
\subsubsection{Terminology and background}
Let $\Delta(\mathcal{A})$ denote the set of all probability distributions over $\mathcal{A}$. For each fixed $b$ the conditional probability $p(a|b)$ defines a probability for each $a \in \mathcal{A}$ and thereby an element $p(A|b)$ in $\Delta(\mathcal{A})$. Define the \textit{convex hull $C(A|B)$ induced by a transition matrix $p(A|B)$} as the set of all the convex combinations of the $p(A|b)$:
\begin{equation}
 C(A|B):=\{p(A) \in \Delta(\mathcal{A})| p(A) = \sum_{b} c_b\; p(A|b)\}
\end{equation} 
here the $c_b,\; b\in \mathcal{B}$ are \textit{convex coefficients}, i.e. for all $b \in \mathcal{B}$ we have $c_b \geq 0$ and $\sum_b c_b =1$. Note that for deterministic transition matrices with full rank (which we will assume in the following) $C(A|B) = \Delta(\mathcal{A})$.

An element $e$ of a convex set $C$ is called an \textit{extreme point} if from $e= \sum_i c_i v_i$ with $v_i \in C, c_i >0$ (note, strictly larger) it follows that $e=v_i$ for all $i$ that are summed over. We denote the set of extreme points of $C(A|B)$ by $E(A|B)$. Note that in general for each extreme point $e \in E(A|B)$ there must exist at least one $b_e \in \mathcal{B}$ such that
\begin{equation}
\label{atleastone}
 e=p^A(A|b_e).
\end{equation} 
Therefore $|\mathcal{B}| \geq |E(A|B)|$. In case of equality $|\mathcal{B}| = |E(A|B)|$ each $p(A|b)$ must correspond to a different extreme point and we get a one-to-one relationship between $b \in \mathcal{B}$ and extreme points $e \in E(A|B)$:
\begin{equation}
\label{extremepoints}
 p^A(A|b)=e_b \text{ and } e=p^A(A|b_e).
\end{equation} 

For any probability distribution $p(A) \in \Delta(A)$ we also define the set $B_A(p(A))$ of states $b$ with $p(A|b)=p(A)$. Note if $e$ is an extreme point of $C(A|B)$ i.e. $e \in E(A|B)$ then from Eq. \ref{atleastone} we know that $B_A(e)$ is not empty.

In the deterministic case $p^A(a|b) := \delta_{f(b)}(a)$.
The sets $B_A(\delta_{i})$ for each $i \in \mathcal{A}$ then partition $\mathcal{B}$ into $|\mathcal{A}|$ blocks and we have $B_A(\delta_{i}) =f^{-1}(i)$. We also have 
\begin{equation}
\label{detextremepoints}
 \{\delta_i(A)|i\in \mathcal{A}\}=E(A|B).
\end{equation} 

\subsubsection{Sketch of proof}
Now assume 
\begin{itemize}
 \item Bayesian network of Fig. \ref{fig:actorsensor}, with $Y \rightarrow S$ and $Z \rightarrow M$,
 \item the stationary distribution of $\{X_t\}_{t \in I}$ has full support (for all $x\in \mathcal{X}$, $p(x)>0$),
 \item strong interaction closure $I(M',X'|S)=0$,
 \item for each $x \in \mathcal{X}$ we have $\pi^{M}(M|x)$ is an extreme point of $C(M|X)$ (e.g. if $\pi^{M}(M|x)$ is deterministic),
 \item $|\mathcal{S}|=|E(M|X)|=:k$ ($=|\mathcal{M}|$ in the deterministic case)
\end{itemize}
A sketch of the proof is as follows.
\begin{enumerate}
 \item First we show that 
\begin{equation}
\label{eequals}
 E(M|X)=: E(M'|X') = E(M'|X) = E(M'|S).
\end{equation} 
\item Then we show that for each $e \in E(M|X) =E(M'|X) = E(M'|S)$ the underlying dynamics $p(x'|x)$ must map elements of $X_{M'}(e)$ into $X'_{M'}(e)$. Similarly, $\pi^{S\dagger}(x|s)$ must map elements of $S_{M'}(e)$ into $X_{M'}(e)$. 
\item Then we prove that the sets $\{X_{M'}(e)|e \in E(M|X)\}$ and $\{X_{M}(e)|e \in E(M|X)\}$ are both partitions of $\mathcal{X}$ which induce functions $\hat{f}^{M'}$ and $\hat{f}^{M}$. Also that $\Pi^S$ is deterministic. We then define $\hat{g}$. Then if $\Pi^M$ is deterministic $\hat{f}^{M'}$and $\hat{g}$ generate $f^{M'},g$ and Eqs. \ref{xGxsupport} and \ref{xGasupport}.
\end{enumerate}
\subsubsection{Proofs}
\begin{description}
 \item[Ad 1.)] Clearly, if two convex sets coincide, then their sets of extreme points coincide. So show first that
\begin{equation}
 C(M'|S) \subseteq C(M'|X') \;\; \text{and} \;\; C(M'|S) \supseteq C(M'|X')
\end{equation} 
Left inclusion first:
\begin{align}
 p(M'|s) &= \sum_{x',x} \pi^M(M'|x') p(x'|x) \pi^{S\dagger}(x|s)\\
         &= \sum_{x'} \pi^M(M'|x') p(x'|s).
\end{align}
Where we only needed the Bayesian network structure of Fig. \ref{fig:actorsensor}. So each $p(M'|s)$ is a convex combination with coefficients $p(x'|s)$ of the distributions $\pi^M(M'|x')$ which span $C(M'|X')$.

Right inclusion:
\begin{align}
 p(M'|x') &= \sum_{a,x} \frac{p(M',x',x,s)}{p(x')}\\
	  &= \sum_{a,x} \frac{p(M'|s) p(x',x|s)\;p(s)}{p(x')}\\
	   &=\sum_{a} p(M'|s) p(s|x').
\end{align}
Where for the step from the first to second line we used
\begin{equation}
 p(m',x',x|s) = p(m'|s)\;p(x',x|s)
\end{equation} 
which follows directly from Eq. \ref{doublecmi} which states:
\begin{equation}
 I(M':X',X|S)=0.
\end{equation} 
So this time we see that all $p(M'|x')$ are convex combinations of the $p(M'|s)$ which proves the right inclusion. 

The proof of $C(M'|X)=C(M'|X')$ proceeds along the same lines. The sets of extreme points then also coincide i.e. Eq. \ref{eequals} holds. 

\item[Ad 2.)] 
Show that all $x \in X_{M'}(e)$ map into $X'_{M'}(e)$. We have 
\begin{equation}
 e = p(M'|x_e) = \sum_{x'} \pi^{M}(M'|x')\;p(x'|x_e),
\end{equation} 
we see that $e = p(M'|x_e)$ is a convex combination of $\pi^{M}(M'|x')$ with convex coefficients $p(x'|x_e)$. But the only convex combinations that result in an extreme point have positive coefficients only for those $\pi^{M}(M'|x')$ with $\pi^{M}(M'|x')=e$ i.e. those $\pi^{M}(M'|x')$ with $x' \in X'_{M'}(e)$, i.e. 
\begin{equation}
\label{xgxweak}
 p(x'|x_e) = \begin{cases} 
0 &  \text{if} \;x' \notin X'_{M'}(e)\\ 
\geq 0 & \text{else},
\end{cases}
\end{equation} 
which proves the condition on $p(X'|X)$. The proof that 
\begin{equation}
\label{xgsweak}
 \pi^{S\dagger}(x|s) = \begin{cases} 
0 &  \text{if} \;x \notin X_{M'}(e_s)\\ 
\geq 0 & \text{else}
\end{cases}
\end{equation}
proceeds along the same line. Notice that each $s \in \mathcal{S}$ is an $s_e$ (Eq. \ref{extremepoints}) and we therefore moved the index in Eq. \ref{xgsweak}.

\item[Ad 3.)] $\{X_{M'}(e)|e \in E(M|X)\}$ is a partition iff a.) for $e_1 \neq e_2 \in E(M|X)$  $X_{M'}(e_1)$ and $X_{M'}(e_2)$ are disjoint and b.) for all $x \in \mathcal{X}$ there exists $e \in E(M|X)$ with $e \in X_{M'}(e)$. Note a.) is true by construction. We show b.). Take an arbitrary $x^* \in \mathcal{X}$.
Notice that there exists $s^* \in \mathcal{S}$ with $\pi^S(s^*|x^*)>0$ because $\Pi^S$ has full rank. But then via definition (Eq.\ref{binverse}) and using that $p(X)$ has full support we get  $\pi^{S\dagger}(x^*|s^*) > 0$. But Eq. \ref{xgsweak} tells us that then $x^* \in X_{M'}(e)$ for some unique $e$. This means for every $x^* \in \mathcal{X}$ there is $e_{x^*}$ with $x^* \in X_{M'}(e_{x^*})$. This proofs b.) and allows us to define a function $\hat{f}^{M'}:\mathcal{X}\rightarrow E(M|X)$ via $\hat{f}^{M'}(x^*)=e_{x^*}$. Then $(\hat{f}^{M'})^{-1}(e)=X_{M'}(e)$. 

Next show that $\{X_{M}(e)|e \in E(M|X)\}$ is a partition. Recall $X_{M}(e)=X'_{M'}(e)$ because of time independence of the high-level processes. Again disjointness is clear. 
Notice that because the underlying process is positive recurrent (as it has a stationary distribution) there exists $x \in \mathcal{X}$ with $p(x'^*|x)>0$. Then from Eq. \ref{xgxgeneral} there must exists a unique $e$ with $x'^* \in X'_{M'}(e)$. So $\{X_{M}(e)|e \in E(M|X)\}$ is also a partition and we define the function $\hat{f}^M$ analogous to $\hat{f}^{M'}$.
We can now extend Eq. \ref{xgxweak} and get:
\begin{equation}
\label{xgxgeneral}
 p(x'|x) = \begin{cases} 
0 &  \text{if} \;x' \notin X'_{M'}(e_x) = (\hat{f}^M)^{-1}\circ\hat{f}^{M'}(x)\\ 
\geq 0 & \text{else}.
\end{cases}
\end{equation} 
Now show that $\Pi^S$ is deterministic. Let $s_1 \neq s_2$ and $\pi^S(s_1|x),\pi^S(s_2|x) >0$. This implies $\pi^{S\dagger}(x|s_1),\pi^{S\dagger}(x|s_2) >0$ and from Eq.\ref{xgsweak} $x \in X_{M'}(e_{s_1})$ and $x \in X_{M'}(e_{s_2})$ which implies (disjointness) $e_{s_1}=e_{s_2}$ which is 
not possible as $|\mathcal{S}| = |E(M|S)|$ (see Eq. \ref{extremepoints}). We then have an associated function $f^S:\mathcal{X}\rightarrow \mathcal{S}$.
%

Define $\hat{g}:= \hat{f}^{M'}\circ (f^S)^{-1}$ (it is bijective). Then 
\begin{equation}
\label{xgsweak2}
 \pi^{S\dagger}(x|s) = \begin{cases} 
0 &  \text{if} \;x \notin X_{M'}(e_s) = (\hat{f}^{M'})^{-1}\circ \hat{g}(s) \\ 
\geq 0 & \text{else}
\end{cases}
\end{equation}

If $\Pi^M$ is deterministic, $e=\delta_i$ ($i \in \mathcal{M}$ see Eq. \ref{detextremepoints}) and we define $f^{M'}$ and $g$ by requiring: if $\hat{f}^{M'}(x)=e=\delta_i$ then $f^{M'}(x):=i$ and if $\hat{g}(s)=e=\delta_i$ then $g(s)=i$. If this is plugged into Eqs. \ref{xgxgeneral} and \ref{xgsweak2} we get Eqs. \ref{xGxsupport} and \ref{xGasupport}. End of proof.
%
\end{description}